\documentclass[conference]{IEEEtran}
\usepackage{amsthm,amssymb,amsmath,subfigure,hyperref,algorithm,algpseudocode,booktabs}
\usepackage{cite}
\usepackage{enumitem}
\usepackage{amsmath,amssymb,amsfonts}
\usepackage{graphicx}
\usepackage{textcomp}
\usepackage{color, soul}
\usepackage[dvipsnames,table,xcdraw]{xcolor}
\usepackage{bm}
\usepackage{subfigure}
\usepackage{siunitx}
\usepackage{hyperref}
\usepackage[nocomma]{optidef}
\newcommand\blfootnote[1]{%
  \begingroup
  \renewcommand\thefootnote{}\footnote{#1}%
  \addtocounter{footnote}{-1}%
  \endgroup
}

\begin{document}

\title{Attack Detection and Localization in Smart Grid with Image-based Deep Learning}

\IEEEaftertitletext{}
\author{\IEEEauthorblockN{\textbf{Mostafa Mohammadpourfard}\IEEEauthorrefmark{1}, \textbf{Istemihan Genc}\IEEEauthorrefmark{1},\\ \textbf{Subhash Lakshminarayana}\IEEEauthorrefmark{2}, \textbf{Charalambos Konstantinou}\IEEEauthorrefmark{3}}
\IEEEauthorblockA{\IEEEauthorrefmark{1}Istanbul Technical University, Istanbul, Turkey}
\IEEEauthorblockA{\IEEEauthorrefmark{2}University of Warwick, Coventry, United Kingdom}
\IEEEauthorblockA{\IEEEauthorrefmark{3}KAUST, Thuwal, Saudi Arabia}\thanks{$^{\ddag}$ Email (corresponding author): mohammadpourfard@itu.edu.tr}\vspace{-8mm}
}

\IEEEaftertitletext{}
\maketitle

\begin{abstract}
Smart grid's objective is to enable electricity and information to flow two-way while providing effective, robust, computerized, and decentralized energy delivery. This necessitates the use of state estimation-based techniques and real-time analysis to ensure that effective controls are deployed properly. However, the reliance on communication technologies makes such systems susceptible to sophisticated data integrity attacks imposing serious threats to the overall reliability of smart grid. To detect such attacks, advanced and efficient anomaly detection solutions are needed. In this paper, a two-stage deep learning-based framework is carefully designed by embedding power system's characteristics enabling precise attack detection and localization. First, we encode temporal correlations of the multivariate power system time-series measurements as 2D images using image-based representation approaches such as Gramian Angular Field (GAF) and Recurrence Plot (RP) to obtain the latent data characteristics. These images are then utilized to build a highly reliable and resilient deep Convolutional Neural Network (CNN)-based multi-label classifier capable of learning both low and high level characteristics in the images to detect and discover the exact attack locations without leveraging any prior statistical assumptions. The proposed method is evaluated on the IEEE 57-bus system using real-world load data. Also, a comparative study is carried out. Numerical results indicate that the proposed multi-class cyber-intrusion detection framework outperforms the current conventional and deep learning-based attack detection methods.\blfootnote{This work was supported by TÜBİTAK and European Commission Horizon 2020 Marie Skłodowska-Curie Actions Cofund program (Project Number: 120C080).}   
\end{abstract}


%
\IEEEpeerreviewmaketitle

\section{Introduction}
The envision for smart grid systems is the bidirectional flow of energy and data between power suppliers and customers, resulting in a more efficient, stable, and automated energy network. However, since smart grids use a variety of information and communication technologies to accomplish this goal, they are more susceptible to cyber-attacks associated with significant political, financial, and physical damages \cite{Ref1a}, \cite{Ref1}, \cite{Ref1c}.  Among the recognized cyber-attacks, attacks on the state estimation (SE) have been widely investigated \cite{Ref3, Ref4}. It is critical for the grid operators to have an accurate estimate of system states since other applications of energy management system (EMS), such as automatic generation control, contingency analysis and etc., rely on SE results \cite{Ref5}.



Incorrect measurements may degrade the accuracy of the outcomes of SE and, therefore, these measurements must be discovered and eliminated. This is accomplished via the use of algorithms known as bad data detection (BDD). \cite{Ref5}. In the same manner, false data injection attacks (FDIAs)  can have a negative impact on SE outcomes \cite{Ref6, Ref25}. In this kind of attack, the adversary modifies the grid's states by manipulating the readings of multiple sensors in such a manner that the conventional BDD is unable to recognize them. Therefore, FDIA has the capability to induce random errors in estimates and consequently  the smart grid control algorithms may be misled, potentially leading in disastrous outcomes such as widespread blackouts and economical losses \cite{Ref7}. To deal with FDIAs, several machine learning-based detectors and mitigation methods have been developed. Table I summarizes important contributions from relevant literature.

Inspired by recent developments in image recognition \cite{Ref15}, and the attempts to frame domain-specific issues as image problems \cite{Ref16}, this paper investigates image-base transformations for FDIA detection. To this end, the developed method includes two steps: (1) power systems' multivariate time series measurements and systems states latent statistical features are extracted by mapping them to image-like data using Gramian Angular Field (GAF)  and Recurrence Plot (RP) \cite{Ref17, Ref18}; (2) a  reliable and robust attack detector is developed to provide a mapping connection between the findings of first phase and attack locations (class labels), using a carefully designed Convolutional Neural Network (CNN)-based structure that is recognized for its effectiveness to accurately capture spatial and temporal correlations in image recognition. This means we treat attack detection problem as an image recognition task.

The developed framework discovers new data features that were not previously accessible for 1D power system measurements, allowing for further detection performance enhancement. As shown in Table I, a major part of existing works relies on one-dimensional signals. Moreover, unlike the existing works which mainly focus on binary classification solutions as shown in Table I, the proposed framework formulates the FDIA detection problem as a multi-class classification problem by applying CCN as a multi-label classifier leading to attack detection and positioning at the same time. 
In summary, our main contributions can be summarized as follows:
\begin{enumerate}
    \item We extract additional statistical features of power system measurements and formulate the FDIA detection issue as an image recognition task by encoding the power system time-series data as image-like data using efficient techniques and without necessitating the use of spatial data of meters. This enables us to utilize the capabilities of lately advanced image-based deep learning algorithms and finally increase the detection accuracy. 
    \item We address the problem of multi-class classification issue by carefully designing the network structure of a multi-label CNN-based classifier to detect attacks but also identify the location of attacks bridging the current gap.  
    \item We perform a comparative analysis utilizing real-world load data to verify the efficiency of the developed framework and present a comparison with current approaches.
\end{enumerate}

\begin{table}[t]
\caption{Summary of related work.}
\vspace{-2mm}
\label{table}
\setlength\tabcolsep{2pt}
\renewcommand{\arraystretch}{1.3}

\begin{tabular*}{\columnwidth}{@{\extracolsep{\fill}}p{1.2cm}p{7.1cm}}

\toprule
Ref&  Main idea \\
\midrule
\cite{Ref3, Ref5, Ref8, Ref9}&   Machine learning algorithms including MLP, random forest, SVM, KNN, etc., are used for binary classification of falsified measurements and normal ones.  \\ 
\cite{Ref10} & A deep belief network coupled with a Gaussian-Bernoulli deep Boltzmann machine is developed as a binary classifier. \\
\cite{Ref4} & The proposed attack detection model utilizes a Deep Neural Network combined with a Decision Tree as a binary classifier, to distinguish attacks from normal samples. \\  
\cite{Ref11,Ref12}& A deep recurrent neural network framework is used for classifying manipulated measurements as attack and normal ones.\\
\cite{Ref13}  & Autoencoder is integrated into an advanced generative adversarial network architecture to spot abnormalities under FDIAs as a binary classifier. \\  
\cite{Ref14} &A binary classification method utilizing the randomized trees and kernel principal component analysis is proposed. \\
\bottomrule
\end{tabular*}
\vspace{-0.3cm}
\end{table}





The remainder of the paper is as follows. The background on FDIAs is discussed in Section II. Section III discusses the utilized time-series transformations employed to encode time-series to images and also the proposed deep learning-based  framework. The numerical results are presented in Section IV. The conclusion of the paper is presented in Section V.

\section{Problem Setting}
The state of a power system is denoted by the magnitudes of the bus voltages $\textbf{V}=[v_{1}, v_{2},\cdots,v_{n}] \in \mathbb{R}  ^{n} $ and the angles $\boldsymbol{\theta}=[\theta_{1}, \theta_{2},\cdots,\theta_{n}] \in \mathbb{R} ^{n}$, where $n$ represents the number of buses. Let $\textbf{z}=[z_{1}, z_{2},\cdots,z_{m}]^T \in \mathbb{R}  ^{m}$ denotes the vector of measurement, $\textbf{x}=[\theta_{1}, \theta_{2},\cdots,\theta_{n}]^T \in \mathbb{R}  ^{n} $ denotes the states variables vector, and  $ \textbf{e}=[e_{1}, e_{2},\cdots,e_{m}]^T \in \mathbb{R} ^{m} $ is the measurement error vector. The following describes the AC measurement model \cite{Ref19}:
\begin{equation}
\textbf{z} = \textbf{h}(\textbf{x}) + \textbf{e}.
\end{equation}

For the purpose of evaluating the FDIA, we use the generally used DC model \cite{Ref3, Ref5, Ref8, Ref10} derived by linearizing the AC model where a matrix $\textbf{H}$  describes the connection between a set of $m$ meter readings and a set of $n$ state variables. In general, $\textbf{H}\in \mathbb{R}^{m \times n}$ is the linear measurement function. In DC SE, $\textbf{x}$ is made up of voltage angles:
\begin{equation}
\textbf{z} = \textbf{Hx} + \textbf{e}.
\end{equation}
The state estimate is often obtained via a weighted least squares estimation as ${\widehat{\textbf{x}}} = (\textbf{H}^T \textbf{W}^{-1} \textbf{H})^{-1} \textbf{H}^T \textbf{W}^{-1}\textbf{z}$, where $ \textbf{W}$ is the covariance matrix. To identify inaccurate data in modern power systems, residual analysis is used. The residual $\|r\|_{2}=\|{\textbf{z}-\textbf{H}\widehat{\textbf{x}}}\|_{2}$ is calculated first, and if it is larger than a predefined threshold value $\tau$, it is assumed that an error occurred. However, as shown in \cite{Ref6}, conventional BDD is susceptible to FDIAs. Particularly, an adversary who knows about $\textbf{H}$, can meticulously craft a malicious vector $\textbf{a} = \textbf{Hc}$ to be added to the original measurement, where $\textbf{c} \in \mathbb{R}^{n}$ is a non-zero injected arbitrary error vector into the real estimates of the systems ${\widehat{\textbf{x}}}$ as illustrated in Fig. 1. Running SE with a malicious measurement $\textbf{z}_{a} = \textbf{z} + \textbf{a}$  will result in an erroneous system state $\widehat{\textbf{x}}_{a} = \widehat{\textbf{x}} + \textbf{a}$. However, FDIA will circumvent the BDD since it will not lead a change in $\|r\| _{2}$:

\begin{equation}
\begin{aligned}
& \|r_{a}\|_{2} = \|{\textbf{z}_{a}-\textbf{H}\widehat{\textbf{x}_{a}}}\|_{2} = \|{\textbf{z} + \textbf{a} -\textbf{H}(\widehat{\textbf{x}} + \textbf{c})}\|_{2} =  \\
      & \| \textbf{z} - \textbf{H}\widehat{\textbf{x}}+(\textbf{a}-\textbf{Hc}) \|_{2} = \|{\textbf{z}-\textbf{H}\widehat{\textbf{x}}}\|_{2}  = \|r\|_{2}
\end{aligned}
\end{equation}

\begin{figure}[t]
\centering
{\includegraphics[scale=0.52]{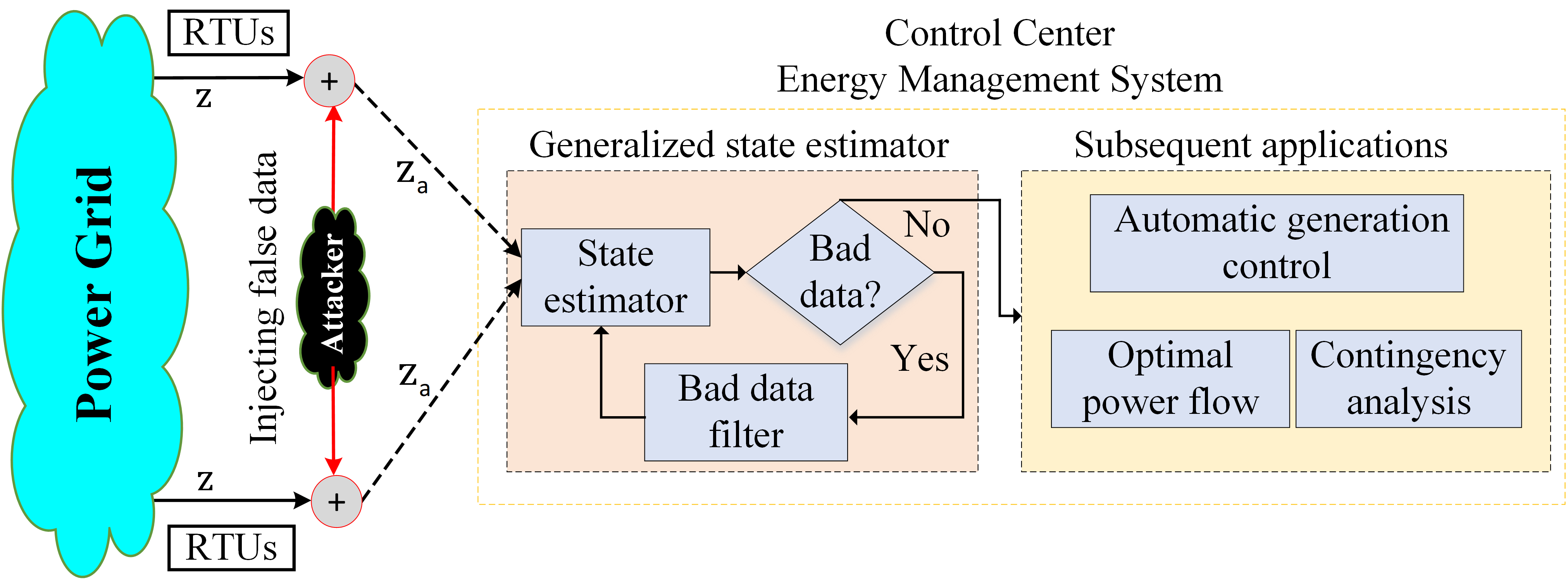} }\\
{\vspace{+0.15cm}
{\footnotesize Fig. 1: Overview of energy management system and FDIA concept.}}
\label{fig:EcUND} 
\end{figure}


\begin{figure*}[htb]
\centering
{\includegraphics[width=\textwidth]{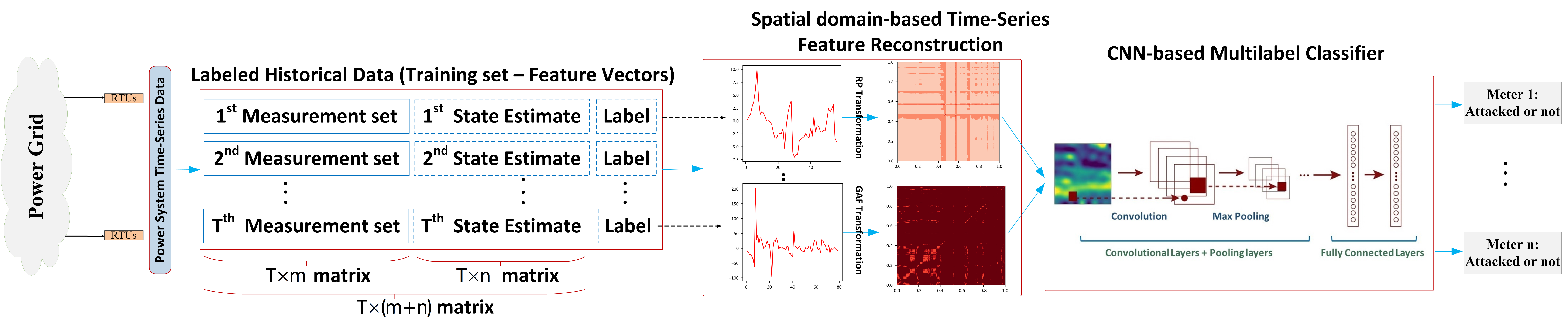} }\\
{\vspace{+0.15cm}
{\footnotesize Fig. 2:  Overview of the proposed method.}}
\label{fig:EcUND} 
\end{figure*}
\section{Proposed Methodology}\label{SecIII}
As previously stated, majority of the current work on FDIAs is devoted to identifying whether an attack occurs or not. However, in reality, apart from attack detection, it is essential to identify the source of the attack (attack location) so that appropriate countermeasure may be deployed quickly. From the view of machine learning, this means they are developed as a binary classifier only showing whether the received measurement is a manipulated one or a normal one. Locating an attack can be seen  a multi-class classification task that continues to generate considerable academic interest owing to its complexity and broad applicability. 
To address the problem, a two-stage deep learning-based mechanism is proposed. We begin by encoding power system time series measurements as image-like structure in order to leverage not only the time series' intra-class correlations but also their inter-class correlations to uncover new data characteristics, and thus, improving attack locating accuracy. To this end, two spatial domain techniques of restructuring measurement data into images namely GAF and RP are used. Since every form of image transformation method encodes data in unique patterns and emphasize various characteristics, it is essential to use different distinct methods for analyzing the image data obtained from power system time-series data to determine the optimal technique for FDIA multilabel classification method. Additionally, this transformation enables us to exploit newly emerging image-based deep learning methods to approach FDIA detection and localization as an image classification problem.

In the second step, we meticulously build a reliable and resilient multilabel CNN-based classifier to accurately capture spatial and temporal correlations and create an end-to-end mapping connection between produced image-based data and the attack location and type which makes our solution robust than current binary classification solutions. The proposed mechanism is depicted in Fig.  $2$.

\subsection{Formation of GAF}
Images obtained from GAF represent data in a polar coordinate system. The dataset $XG_R=\{xg_1,xg_2,xg_3,…,xg_l,…,xg_n\}\in \mathbb{R}$ is a time series consisting of $n$ samples. Values of the $XG_R$ are re-scaled between a range $[0,1]$ and the new vector is named $XG_S$. Then, the rescaled time-series data is encoded according to the following equation \cite{Ref17}:
\begin{equation}
\begin{cases}
   \begin{array}{c}
\phi=\arccos \left(x_{lg 0}\right),\\
r=\frac{t_{l}}{N}
\end{array}
\end{cases}
\end{equation}
where $x_{lg 0}$ corresponds the $l_\text{th}$ observation of the $XG_S$, $N$ and $t_l$ are the fixed factor and timestamp, respectively. In order to identify the temporal correlation of the encoded image within different time intervals, the sum/difference between each point of the image can be used. In this paper, the summation method is used for the GAF based on the following equations:
\begin{equation}
GAF = cos(\O_{l} + \O_{k})
\end{equation}
\begin{equation}
GAF = X^{T}_{S}.X_{S}-\sqrt{I-{X^{2}_{S}}}^{T}.\sqrt{I-{X^{2}_{S}}},
\end{equation}
where $I$ and $n$ represent the unit row vector and the length of the inertia time series data, respectively.

GAF images generated from power system time-series measurement with and without FDIA are illustrated in Fig. 3 . It is observed from Figs. 3(b) and 3(d) that encoding latent characteristics from time series data into pictures through the GAF transformation technique results in distinct patterns for an attack and a normal sample, thus improving the proposed classifier's performance when comparing raw time series data (Figs.  3(a) and  3(c)).

\begin{figure}[t]
\centering
{\includegraphics[scale=0.6]{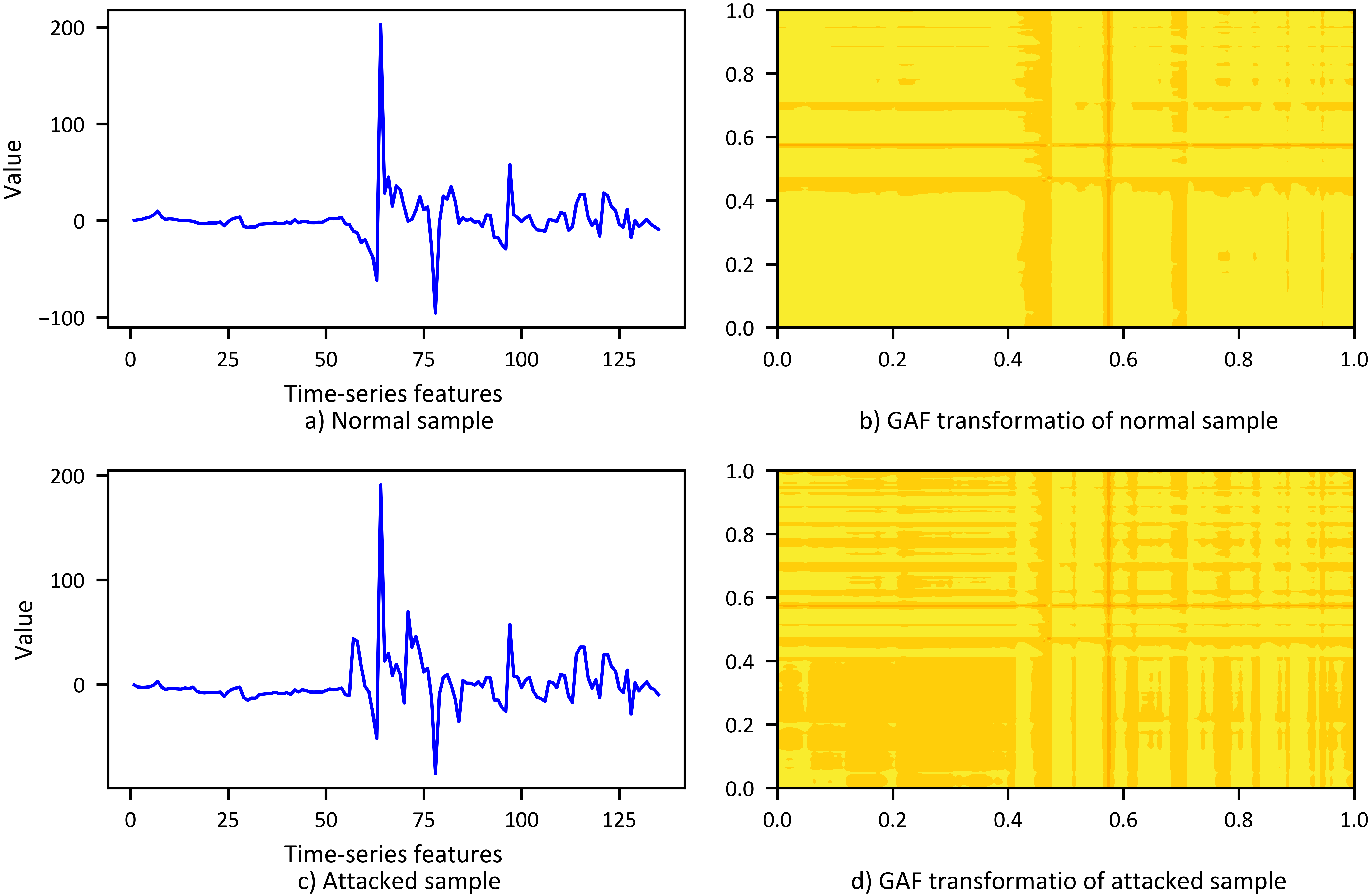} }\\
{\vspace{+0.15cm}
{\footnotesize Fig. 3: GAF-based time-series feature reconstruction.}}
\label{fig:EcUND} 
\end{figure}

\subsection{Formation of RP}
Time series data have some recurrent behaviors such as periodicity and irregular cycles. Visualization of the structure of such data, especially the structure of recurrence of a time series using a tool called RP is possible [5]. The RP is an image indicating the distances between time points of a multivariate time series \cite{Ref18}. For a multivariate time-series such as $q(t) \in \mathbb{R}$, the PR is expressed as follows:
\begin{equation}
   RP=\Theta (\varepsilon -\mid \mid{{q}_{( i )}}-{{q}_{( j )}}  \mid \mid),
\end{equation}
where $\varepsilon$ is the threshold and $\Theta$ is a function called heaviside. The images generated by RP is shown in Fig. 4. As it can be seen, the image transformation of a normal measurement and an attack measurement has distinct characteristics and patterns, even though the distributions of raw data are similar.

\begin{figure}[t]
\centering
{\includegraphics[scale=0.6]{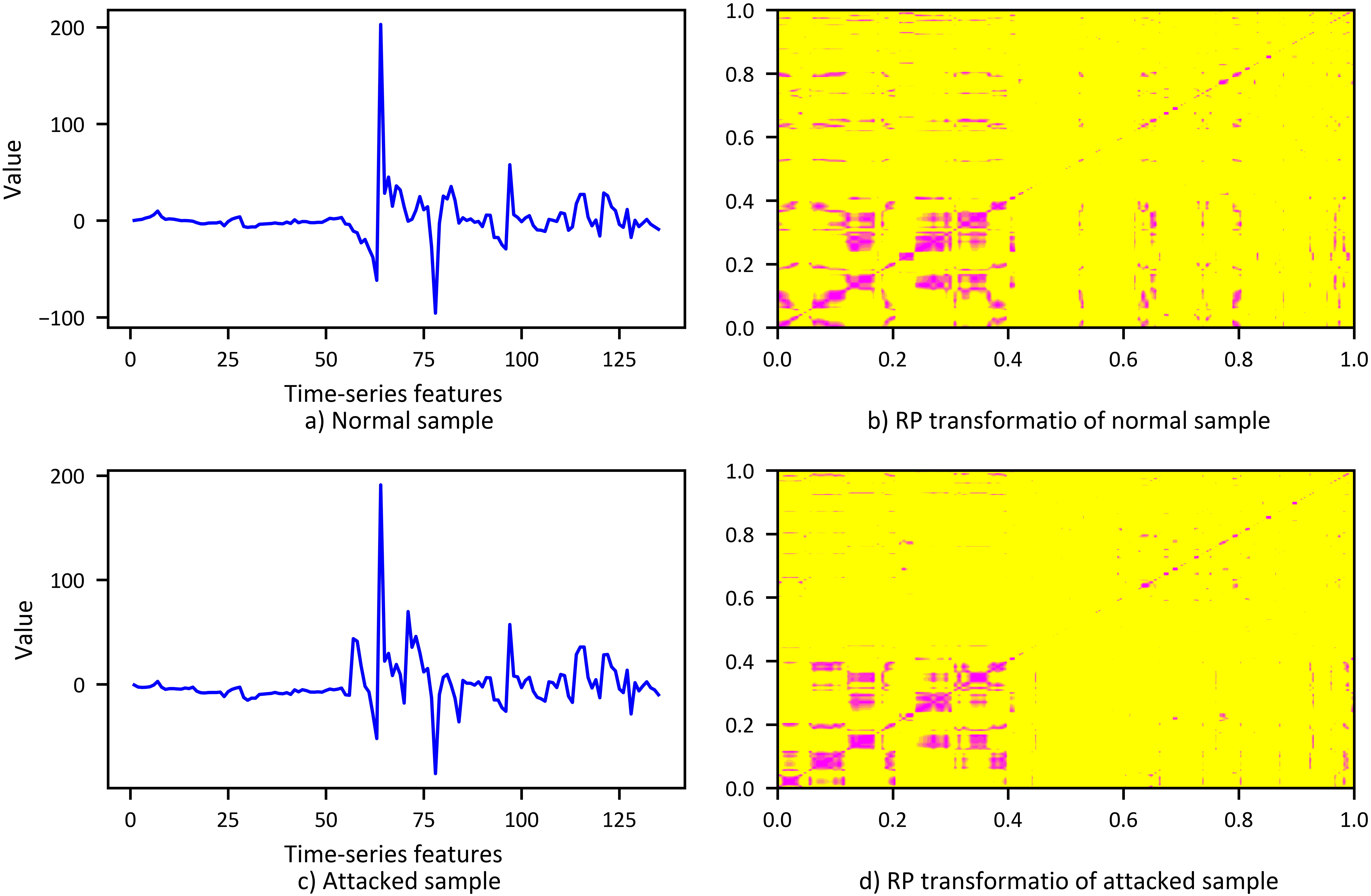} }\\
{\vspace{+0.15cm}
{\footnotesize Fig. 4:  RP-based time-series feature reconstruction.}}
\label{fig:EcUND} 
\vspace{-0.2cm}
\end{figure}
\subsection{Convolutional Neural Network (CNN)}
CNN is designed as one of the deep learning applications to solve classification problems, image or video processing, and data processing of multiple arrays. The CNN has some advantages over other deep learning methods, which adds to its performance \cite{Ref20}. CNN is composed of four layers: convolution layer, pooling layer, fully-connected layers, and finally classification, shown in Fig.  5. The convolution and pooling layers form the first few steps of CNN. Each convolution layer is organized so that each feature space in it has some filters to extract features. The extracted features by each filter are collected by the pooling layer in the same layer during these steps. Pooling operations occur using max-pooling to prevent overfitting. Max pooling is the type of pooling that embraces most of the extracted features. 
\begin{figure}[t]
\centering
{\includegraphics[scale=0.44]{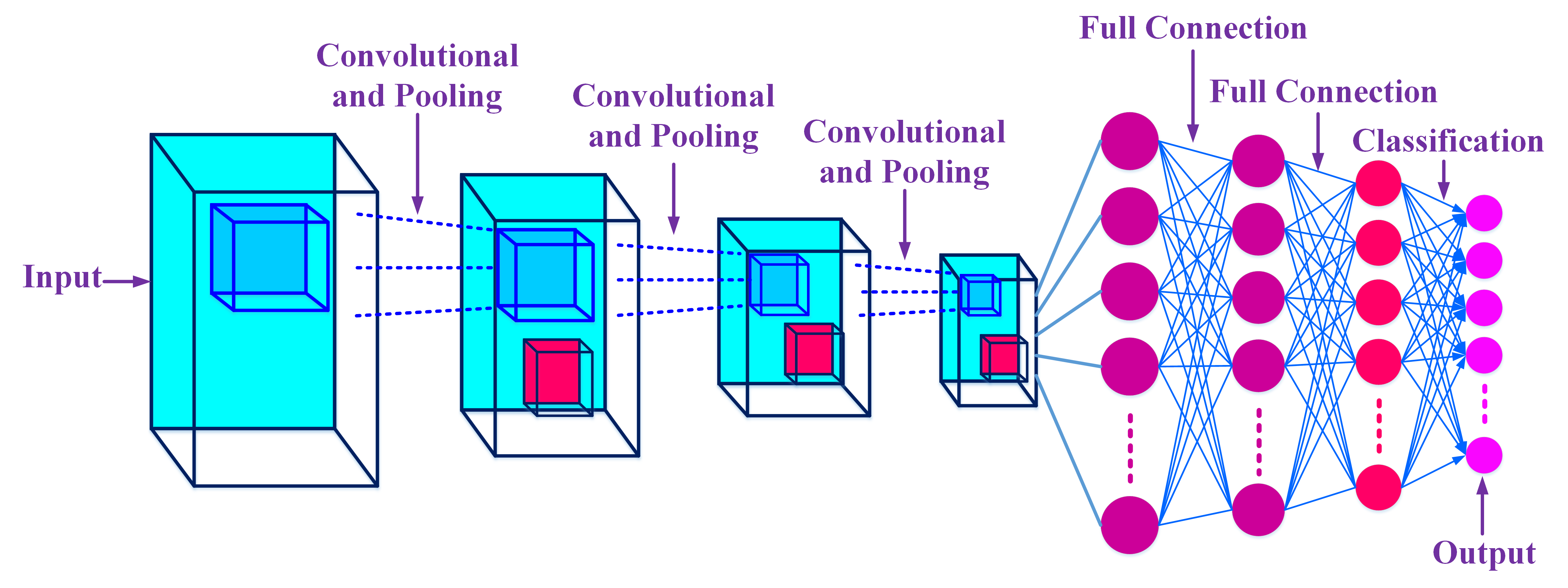} }\\
{\vspace{+0.15cm}
{\footnotesize Fig. 5: The main architecture of CNN.}}
\label{fig:EcUND} 
\end{figure}
The output of each pooling layer is transmitted by rectified linear unit (ReLU) as input to the next convolution layer. Batch normalization is used to avoid the internal covariate shift after the activation layer. These steps continue until to reach the fully connected layers. The feature vector prediction at the network output is the task of the fully connected layers, and finally, a Softmax function is used to classify extracted features in the last layer of fully connected layers. In the convolutional step, a ReLU is used as follows:

\begin{equation}
    C_{r}^{n}=R e L U({\sum_{m} v_{r-1}^{m}\text{*}w_{r}^{n}+b_{r}^{n}}),
\end{equation}
where $C_r^n$ shows the response of $n^{th}$ filter in convolutional layer $r, v_{r-1}^{m}$ denotes the $m^{th}$ output of the last layer $r-1$, operator $\ast$ shows the convolution, and $w_{r}^{n}$ indicates the $n^{th}$ filter kernel of the running layer $r$, and $b_{r}^{n}$ depicts the bias. The Max pooling follows \cite{Ref21}:
\begin{equation}
h^{n}_{r+1}=\max  C^{n}_{r}(t), \qquad (i-1)l+1\leq t \leq il,
\end{equation}
where $h^{n}_{r+1}$ shows the response of pooling layer, $l$ shows the length of local area for pooling, $t$ is the number of collected features by the Max pooling, and $i$ is the number of input features to the pooling layer. Finally, a Softmax function is used as follows:
\begin{equation}
\small
O_j= \begin{bmatrix}  
P(y=1)|x;\vartheta \\
P(y=2)|x;\vartheta \\
\dots \\
P(y=k)|x;\vartheta
\end{bmatrix} 
=\frac{1}{\sum_{j=1}^{k} exp(\vartheta^{j}x)}
\begin{bmatrix}  
exp(\vartheta^{1}x)  \\
exp(\vartheta^{2}x)\\
\dots \\
exp(\vartheta^{k}x)
\end{bmatrix},
\end{equation}
where $k$ shows the number of categories and $\vartheta^jx$ are the factors of the classification layer.

\section{Numerical Results}\label{SecIV}

\subsection{Data Preparation}
IEEE $57$-bus system is used to complete the simulations using MATPOWER \cite{Ref22}. As illustrated in Fig.  6, this system contains 7 generators and 80 branches. Real-world load data were obtained from the New York Independent System Operator (NYISO)  to improve the accuracy of subsequent simulations \cite{Ref23, Ref23a}. The data covers the first week of the January $2020$. Due to the fact that the NYISO load profiles are only 11, sampled every 5 minutes, all profiles were duplicated and re-scaled to fit the used test system and ensure solvability of the SE. For each day, there are 288 load values. This gives us the data values of $T = 2045$ that corresponds to the normal operation points for each area (i.e., without FDIAs). The noise added to measurements is assumed to be Gaussian with a mean of zero and a standard deviation of $2\%$.

\begin{figure}[b]
\centering
{\includegraphics[height=0.35\textheight,width=0.42\textwidth]{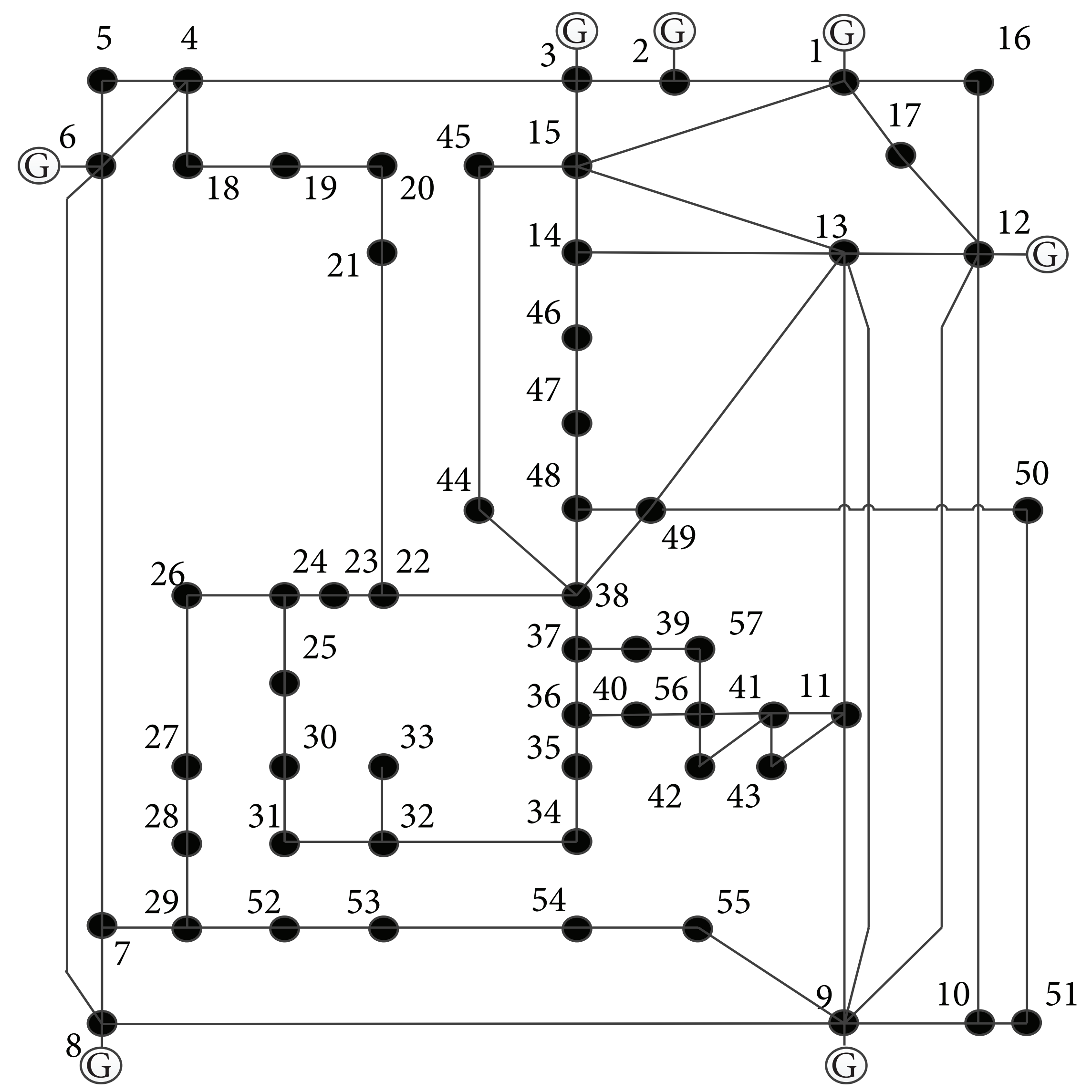} }\\
{\vspace{+0.15cm}
{\footnotesize Fig. 6: IEEE 57-bus test system.}}
\label{fig:EcUND} 
\end{figure}
\subsection{Attack Simulation}
False data are injected into 13 randomly selected buses (\#2, 6, 10,14, 19, 25, 31, 35, 38, 43, 47, 51, and 57). Therefore, we have 14-class labels (13 attack classes and 1 normal class). In contrast to the majority of currently available techniques, which are primarily focused on two classes of labels (attack and normal) and binary classification, we have a multi-label dataset and can determine the location of attacks. For each attacked system state, we inject two distinct FDIAs that are $90\%$ and $110\%$ of their true values, respectively. Active power measurements are  selected as measurements. This means there are 80 measurement features. These measurements, along with their corresponding state variables except the slack bus, are chosen as inputs of the proposed framework. An overview of the utilized dataset is presented in Table 2. System operation is assumed to be normal until the $7_{\text{th}}$ day. This means that the normal data for the first 6 days are  accessible, and the last day's measurements are attacked.

\begin{table}[t]
\caption{Overview of the used dataset.}
\label{table}
\centering
\setlength\tabcolsep{1.8pt}

\begin{tabular*}{\columnwidth}{@{\extracolsep{\fill}}p{1.68cm}p{1.68cm}p{1.68cm}p{1.68cm}p{1.68cm}}
\toprule
  No. of features& No. of normal samples &  No. of attack samples per class label& No. of all samples& No. of class labels \\
\midrule
136& 2045 & 576  & 9533&14\\
\bottomrule
\end{tabular*}	
\end{table}
\subsection{Performance Evaluation Metrics}
In our experiments, $F_{1}$ score is utilized as performance evaluation metric. $F_{1}$ is defined as:
\begin{equation}
F_{1}= (\frac{2\times Precision\times Recall}{Precision + Recall}), 
\end{equation}
where precision and recall are defined as:
\begin{equation}
P_{r}= (\frac{TP}{TP + FP}) \quad \quad R_{e}= (\frac{TP}{TP + FN}),
\end{equation}
where true positive (TP), false positive (FP), and false negative (FN) are defined as the likelihood that a attacked bus will be classified correctly, a normal sample sample  is labeled as attacked, and a compromised location is labeled as normal, respectively.
\subsection{Image Transformation}
The generated attack and normal samples are transformed into GAF and RF images using the explained formulations in Section III. This resulted in two datasets to be fed into a carefully designed CNN-based multi-label classifier for solving our mutli-class classification problem and comparing the results.
\subsection{Proposed multilabel CNN Architecture}
The proposed multilabel CNN-based classifier is trained on the 2D generated images and output the multi-label image classifications. The network is composed of several convolutional layers (Conv), one flattening layer, and one fully connected hidden layer. The convolutional parameters are fine-tuned using Conv1 and Conv2, and as a result, both layers have the same amount of feature maps, with a kernel size of 32 and ReLU activation function. Conv3 and Conv4 serve as tuning features, and therefore, the parameter settings are the same, and the kernel size is 64, and supplied with the activation function of ReLU. Conv5 is critical in rebuilding feature maps and converting them to channel output. As a result, the Conv5 kernel size is set to 128. Additionally, to address the issue of overfitting, the dropout = 0.25 is  utilized with the last convolutional layer. Finally, the designed structure completes the multi-class classification task through receiving flattened feature maps by a fully connected layer with SoftMax function. The CNN was trained by the Adam optimization algorithm, batch-size of 128, epoch number of 200, and loss function of \emph{categorical-crossentropy}. All two image datasets are randomly splitted into two distinct subsets for training ($70\%$) and testing ($30\%$). 
\subsection{Results}
The findings of the proposed framework are summarized in Table III. Additionally, we compare the proposed two-stage framework with the well known state-of-the-art techniques, including support vector machine (SVM), multi-layer perceptron (MLP)-based deep learning methodology which its architecture is similar to the CNN, k-nearest neighbor (k-NN), and random forest. It is noteworthy that to calibrate the hyperparameters and obtain the proper sets that assigns the highest F1-Score, the grid search method is utilized \cite{Ref24}. For example, for SVM and MLP the best values for important hyperparameters were: \{kernel= $rbf$, C=1000, gamma =0.01\} and \{number of hidden layers= 2, number of neurons = (64,128), batch size=128, activation= ReLU, optimizer= Adam\}, respectively. To provide a fair comparison, we train and evaluate all algorithms on the same datasets.

Compared to existing established techniques, substantial performance improvement is obtained in detecting FDIA by the proposed method. According to Table III, the model that achieved the best results is the proposed RP-based robust constructed CNN-based multi-label classifier. Additionally, the developed RP-based method's performance is described using the confusion matrix shown in Fig.  7.

\begin{table}[t]
\caption{Summary of results.}
\label{table}
\centering
\setlength\tabcolsep{2pt}

\begin{tabular*}{\columnwidth}{@{\extracolsep{\fill}}llll}
\toprule
  Approach& Precision &  Recall& $F_{1}$ \\
\midrule
\textbf{RP-CNN}& \textbf{0.99} & \textbf{0.99}  & \textbf{0.99}\\
GAF-CNN& 0.99 & 0.97  & 0.98\\
SVM& 0.9 & 0.9  & 0.9\\
MLP& 0.92 & 0.91  & 0.91\\
k-NN& 0.85 & 0.76  & 0.8\\
Random Forest& 0.89 & 0.88  & 0.88\\
\bottomrule
\end{tabular*}	
\end{table}
\begin{figure}[t]
\centering
{\includegraphics[scale=0.5]{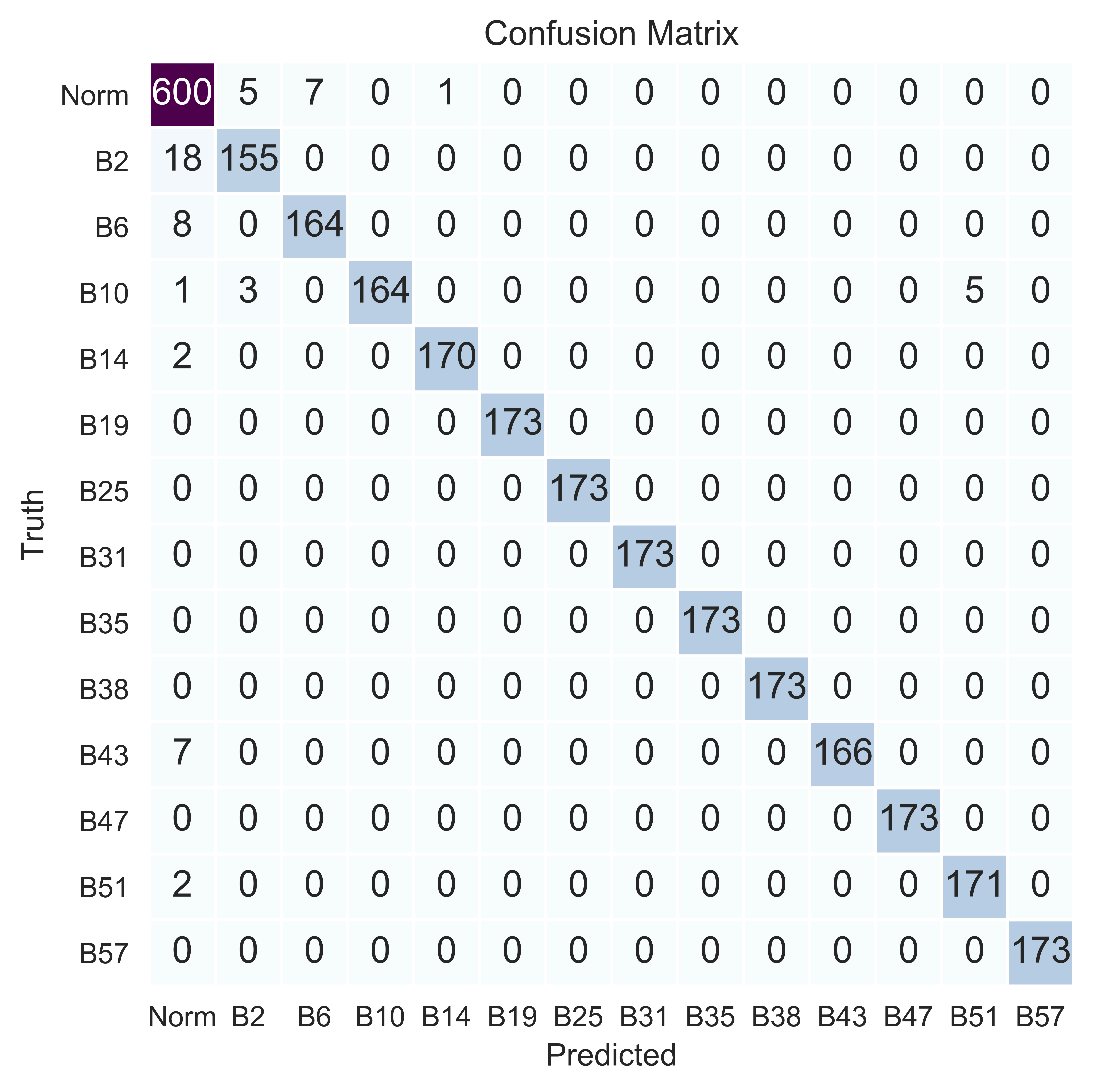} }\\
{\vspace{+0.15cm}
{\footnotesize Fig. 7: Confusion matrix of the proposed method for test dataset. }}
\label{fig:EcUND} 
\end{figure}
During experiments, it is is observed that the state-of-the-art techniques show random behavior in localizing some of the attacks. For example, Table IV shows the results for locating attacks on the buses 2, 6, and 43 for the proposed method and MLP and SVM, which showed a better accuracy compared the the other methods. As one can see, the proposed method is robust in locating attacks on different buses while the current methods behave randomly, making them unreliable to be used in the real-world power grid applications. 

\begin{table}[t]
\caption{Behavior of other techniques in attack location.}
\label{table}
\centering
\setlength\tabcolsep{2pt}

\begin{tabular*}{\columnwidth}{@{\extracolsep{\fill}}llll}
\toprule
  Approach& Bus 2 &  Bus 6& Bus 43 \\
\midrule
RP-CNN& 0.92 & 0.96  & 0.98\\
SVM& 0.73 & 0.88  & 0.81\\
MLP& 0.56 & 0.85  & 0.87\\
\bottomrule
\end{tabular*}	
\vspace{-0.2 cm}
\end{table}

\section{Conclusion}\label{SecVII}
In this paper, we have developed a two-stage learning-based approach based on spatial domain techniques and image-based deep learning techniques for detecting and locating FDIAs to enhance the cybersecurity of modern power grids. In particular, we formulate the FDIA detection and localization problem as a multi-label classification problem and finally treat it as a image recognition task. A substantial performance increase is obtained compared to the state-of-the-art detectors by the developed image-based robust designed CNN-based multi-class classifier. Moreover, the framework shows robustness in locating FDIAs.


%
%



%

\end{document}